\documentclass[conference]{IEEEtran}
\IEEEoverridecommandlockouts
\usepackage{cite}
\usepackage{amsmath,amssymb,amsfonts}
\usepackage{algorithmic}
\usepackage{graphicx}
\usepackage{subfig}
\usepackage{textcomp}
\usepackage{xcolor}
\def\BibTeX{{\rm B\kern-.05em{\sc i\kern-.025em b}\kern-.08em
    T\kern-.1667em\lower.7ex\hbox{E}\kern-.125emX}}
\begin{document}

\title{RF Chain-Free mmWave Transmission: Modeling and Experimental Verification
}
\author{\IEEEauthorblockN{
M. Yaser {Yağan}\IEEEauthorrefmark{1}\IEEEauthorrefmark{2},   
İbrahim {Hökelek}\IEEEauthorrefmark{1},
Ali E. {Pusane}\IEEEauthorrefmark{2},
Ali {Görçin}\IEEEauthorrefmark{1}\IEEEauthorrefmark{3}
}                                     
\IEEEauthorblockA{\IEEEauthorrefmark{1}
Communications and Signal Processing Research (HİSAR) Laboratory, T{\"{U}}B{\.{I}}TAK B{\.{I}}LGEM, Kocaeli, Türkiye}
\IEEEauthorblockA{\IEEEauthorrefmark{2}
Department of Electrical and Electronics Engineering, {Boğaziçi} University, İstanbul, Türkiye}
\IEEEauthorblockA{\IEEEauthorrefmark{3}Department of Electronics and Communication Engineering, Istanbul Technical University, {\.{I}}stanbul, Türkiye}
\IEEEauthorblockA{ \emph{yaser.yagan@tubitak.gov.tr, ibrahim.hokelek@tubitak.gov.tr,}}
\IEEEauthorblockA{ \emph{ali.pusane@boun.edu.tr, aligorcin@itu.edu.tr} }
}

\maketitle

\begin{abstract}
The utilization of millimeter wave frequency bands is expected to become prevalent in the following communication systems. However, generating and transmitting communication signals over these frequencies is not as straightforward as in sub-6 GHz frequencies due to complex transceiver structures. As an alternative to conventional transmitter architectures, this paper investigates the implementation of time-modulated arrays to effectively modulate and transmit high-quality communication signals at millimeter wave frequencies. By exploiting the array structures and analog beamformers, which are the fundamental components of millimeter wave transmitters, secure and low-cost transmission can be achieved. Though, harmonics of theoretically infinite bandwidth arise as a fundamental problem in this approach. Thus, this paper presents a frequency analysis tool for the time-modulated arrays with hardware impairments and shows how controlling the sampling period can reduce the harmonics. Furthermore, the derived results are experimentally verified at 25 GHz with two important remarks. First, the phase error of received signals can be reduced by 32\% using the proposed architecture. Second, the harmonics can be significantly suppressed by the correct choice of sampling period for the given hardware. 
\end{abstract}


\section{Introduction}
Utilizing antenna arrays and large surfaces for modulating carriers has been a significant research topic that has drawn recent attention with 6G \cite{RISmodulation1,RISmodulation2,RISmodulation3}. Time-modulated arrays (TMAs) \cite{TMArev} are considered a potential solution that provides low-cost modulation schemes with physical layer security \cite{secure1,secure2}. It has been treated from two perspectives: physical layer security where spatial signal processing is exploited to transmit information in specific directions while ensuring high secrecy rates in other directions, and low-cost transmission where the number of radio frequency (RF) chains in the transmitter is reduced. 

The concept of direct modulation through antenna arrays has been widely studied in the antennas and propagation research society. The periodic pulses applied to individual antenna elements enable controlling the radiation of the harmonic frequencies that arise from this modulation, which is known as sidelobe radiation \cite{TMA2}. Applying pulses directly to antennas fed with RF signals requires very fast RF switches. Thus, TMAs have attracted more focus recently as the fabrication and availability of such components increased. Besides communication, TMAs have various applications in radar and wireless power transfer. 

The control of the harmonic levels in TMAs was a major problem that has been studied in the literature \cite{TMA1}. The authors in \cite{TMAPS} investigated pulse shaping in TMAs and showed its effect in suppressing the harmonics. Many other works \cite{TMA2,TMA3,TMA4,TMA5} considered harmonic suppression by either examining the pulse shape effect or controlling the pulse sequences. As a result, up to 25 dB and 30 dB of harmonic suppression have been reported in \cite{TMA4} and \cite{TMA5}, respectively.

From an implementation perspective, two issues have not been addressed in the literature: The first one is the modulated signal bandwidth and harmonic frequencies, where all the works mentioned considered the harmonic index independent from its real value, which in turn has a significant effect on suppression. The second problem is the response of the modulating circuit. The authors in most works developed their solution with the assumption that each modulation symbol is presented by a single pulse, and the pulse shape is realized by the modulating circuit. However, the work of \cite{TMA3} has considered the problem of having a symbol period different from the pulse period. Accordingly, two restrictions on the signal bandwidth and the pulse period were introduced. 

Aside from these aspects, generating communication signals at millimeter wave (mmWave) frequency has the high phase noise challenge resulting from mixing at high frequencies \cite{phnoise3,phnoise4}, which increases the RF chain components' cost significantly to compensate for this effect. As mmWave communications proposed with the early 5G preliminary works, the problem of phase noise has been highlighted, and more efforts were put into solving it \cite{phnoise1,phnoise2}. 

These demands motivated us to develop a mmWave transmission model that utilizes analog beamformers to achieve a secure, low-cost, and high-quality modulation. The security of the proposed model comes from the directional modulation algorithm, while the low cost and high quality are the results of RF chain-free hardware that exhibits less phase noise. In this paper, the problem of TMA is treated with a more hardware-oriented focus. By considering that the response of the modulating circuit is unchangeable, an analysis method is derived where the sampling period can be altered with a fixed circuit response time. The proposed model includes a mathematical investigation of hardware impairments and experimental verification. Experimental results show a notable improvement in the phase error of modulated symbols compared to digitally modulated signals. 

The rest of the paper is organized as follows: Section \ref{RFtext} describes the proposed mmWave TMA system architecture, in Section \ref{PStext} the hardware impairments are investigated and a frequency analysis method is proposed, Section \ref{Extext} presents experimental results, and Section \ref{conclusiontext} concludes the paper.

\section{RF Chain-Free Transmitter Model}
\label{RFtext}
\begin{figure}[]
\centerline{\includegraphics[width=\linewidth]{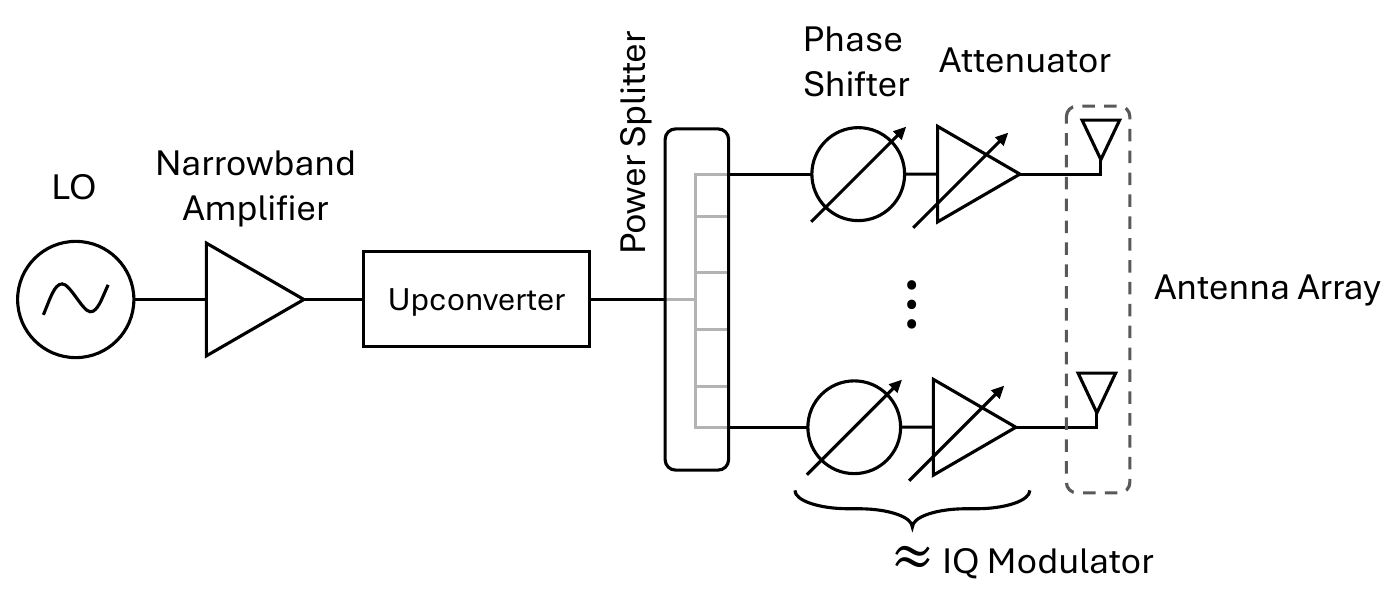}}
\caption{The analyzed transmitter model.}
\label{fig:modelarch}
\end{figure}

An RF chain-free mmWave transmitter model is shown in Fig. \ref{fig:modelarch}. Compared to conventional RF transmitters, the cost reduction is provided by removing the digital-to-analog converters (DACs) and replacing the amplifiers that usually have to satisfy certain bandwidth requirements by narrowband amplifiers that only need to amplify the carrier signal. Besides, in conventional transmitters, the modulated signals are passed through the upconverters, resulting in a higher phase noise compared to the architecture of Fig. \ref{fig:modelarch}, where only the carrier is upconverted. The phase shifters and attenuators used for beamforming are digitally controlled analog components. However, at mmWave frequencies, beamforming is an indispensable enabler, and these components are also utilized in classical mmWave transmitters. By fast and periodic control of the phase shifters and amplifiers, any sequence of IQ samples can be modulated. Beamforming is realized simultaneously by multiplying the modulating samples by constant weights that depend on the controlled antenna element and the desired transmission direction. 
The over-the-air modulation achieved in this model has the advantage of implementing spatial multiplexing to transmit multiple users' data streams, where the weights at each time instance are calculated to deliver the carrier signal at the desired directions in different phase and amplitude levels, and hence transmitting different direction-dependent IQ samples. In contrast to conventional MIMO systems where the number of users can not exceed the number of RF chains in the transmitter, the number of users in this scheme can reach the number of antenna elements. Additionally, the spatial dependency of modulated symbols results in transmitting random or meaningless signals in undesired directions, which increases security. As a result, this transmitter architecture provides a low-cost and secure solution for multi-user transmission.

The array pattern analysis and beamforming-related studies have been conducted in the literature. Furthermore, the focus of this paper is on the bandwidth analysis and the hardware response effects on each single antenna element, independent of the array factor and spatial radiation. 
\section{Hardware Impairments and Pulse Shaping}
\label{PStext}
The modulation scheme described in the previous section assumes applying the calculated complex weights for a sampling period. If the samples per symbol (SPS) rate is 1, then the weights are applied for a symbol period. In practical implementation, two limitations arise: the transition time between two consecutive beamforming weights (the response time) and the maximum achievable switching rate. Fortunately, the former limits the transmission bandwidth, which would be infinite if the transition is ideal. The latter affects the sampling rate and hence the symbol rate. Additionally, as the weight pulses synthesize a periodic sampling process, they produce periodicity in the spectrum, and hence, the bandwidth becomes infinite. While this problem can be solved via filtering, in this section, a solution without filtering is sought as the paper's purpose is cost reduction. 
\subsection{Ideal and Realistic Rectangular Pulse}
\begin{figure}[]
\centering
\subfloat[Ideal and realistic pulses.]{%
\includegraphics[width=0.75\columnwidth]{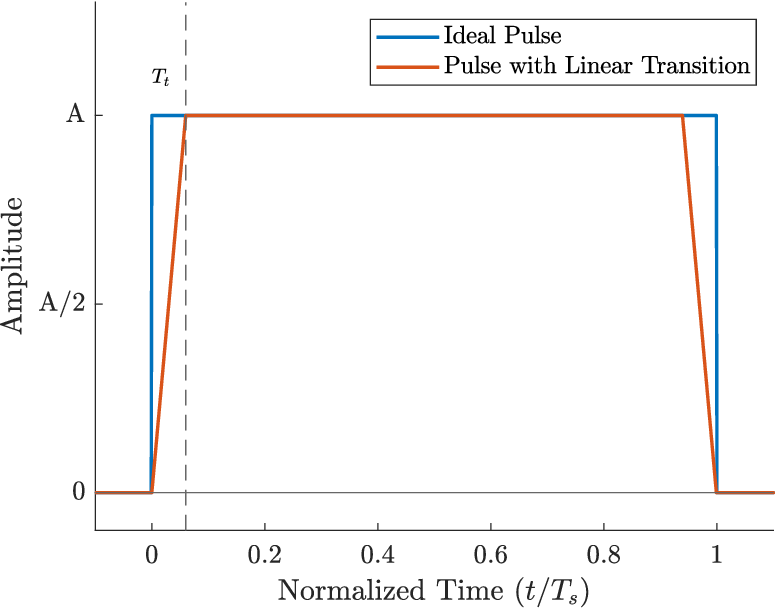}%
} \\
\subfloat[The spectrum of ideal and realistic pulses.]{%
\includegraphics[width=0.75\columnwidth]{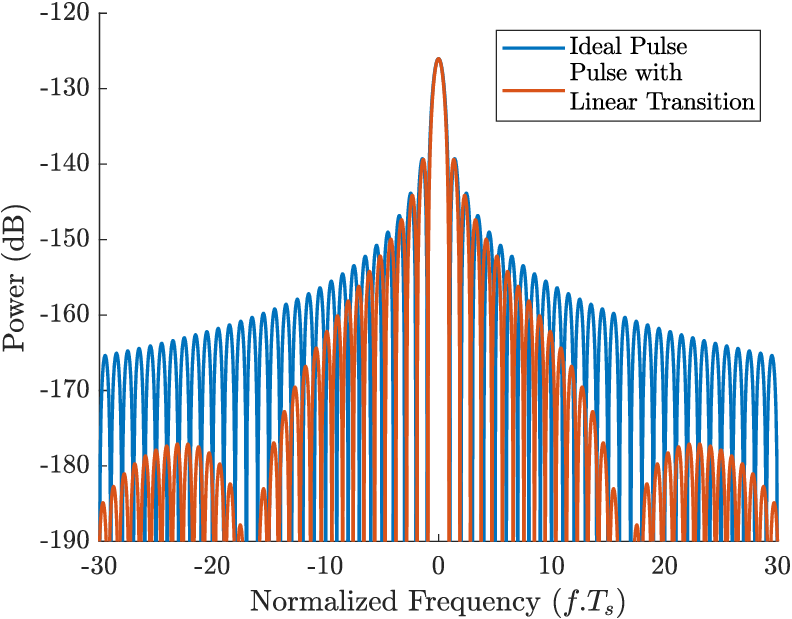}%
} \\ 
\caption{Ideal and realistic pulses with their spectrum.}
\label{fig:ideal}
\end{figure}

We start the analysis by assuming a linear transition between weight pulses. An ideal pulse with $T_s$ duration and $A$ amplitude, as shown in Fig. \ref{fig:ideal} (a), can be expressed as the difference of two step functions $u(t)$ as
\begin{equation}
    x_{\text{ideal}}(t) = A\left[u(t)-u(t-T_s)\right],
\end{equation}
and its Fourier transform is known to be
\begin{equation}
    X_{\text{ideal}}(f) = AT_s sinc \left(T_s \pi f\right) e^{-jT_s\pi f}.
    \label{eq:ideal_f}
\end{equation}
It can be seen in Fig. \ref{fig:ideal} (b) that the ideal pulse occupies a large bandwidth in the spectrum. This is consistent with (\ref{eq:ideal_f}) as the $sinc(.)$ function extends to infinity. A more realistic pulse, with a transition duration of $T_t\ll T_s$ where the amplitude changes linearly as shown in Fig. \ref{fig:ideal} (a), is
\begin{multline}
    x_{\text{real}}(t) = \frac{A}{T_t}[tu(t) 
    -(t-T_t)u(t-T_t) \\
    -(t-T_s+T_t)u(t-T_s+T_t)\\
    +(t-T_s)u(t-T_s)].
\end{multline}
The spectrum of this pulse is then calculated as 
\begin{multline}
    X_{\text{real}}(f) = -A(T_t+T_s) sinc \left(-T_t \pi f\right) \\
    sinc \left(-(T_t+T_s) \pi f\right)e^{-jT_s\pi f}.
    \label{eq:real_f}
\end{multline}
Here, the additional $sinc(.)$ term suppresses the signal after the frequency $1/T_t$, which is relatively high, as seen in Fig. \ref{fig:ideal} (b). As a result, applying the IQ samples directly for a symbol period exhibits spectral inefficiency as the energy is spread over the spectrum. This problem also causes interference issues for frequency duplex multiple access scenarios. While previous works proposed changing the pulse shape directly by controlling the response of the pulse-generating circuit, we investigate a more realistic case where the response characteristics of the modulating circuit, specifically the linear transition and its duration ($T_t$), are accepted as uncontrollable constraints. Thus, the next subsection will investigate increasing the SPS rate. 
\subsection{Pulses with Higher SPS}
\begin{figure}[ht!]
\centering
\subfloat[Trapezoidal pulses with different SPS rates.]{%
\includegraphics[width=0.845\columnwidth]{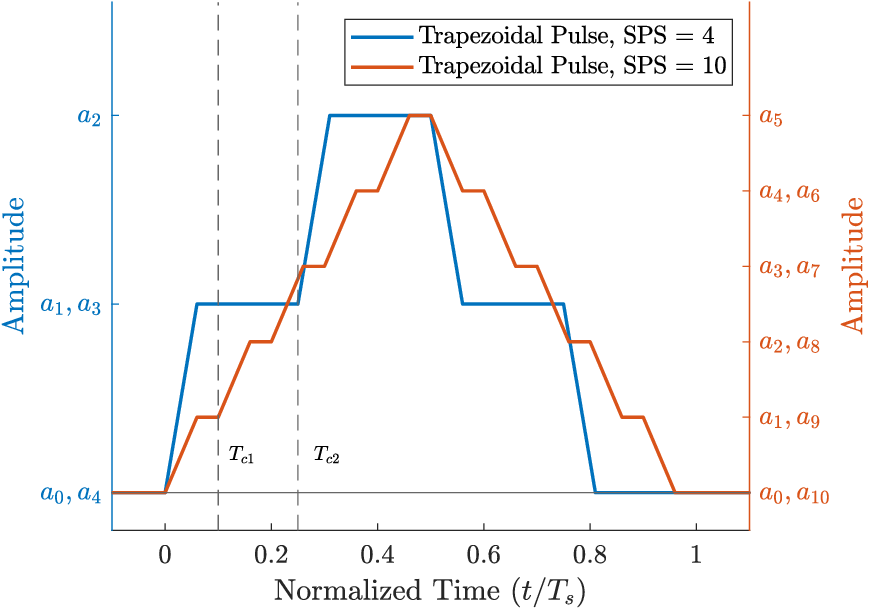}%
} \\
\subfloat[The spectrum of trapezoidal pulses.]{%
\includegraphics[width=0.75\columnwidth]{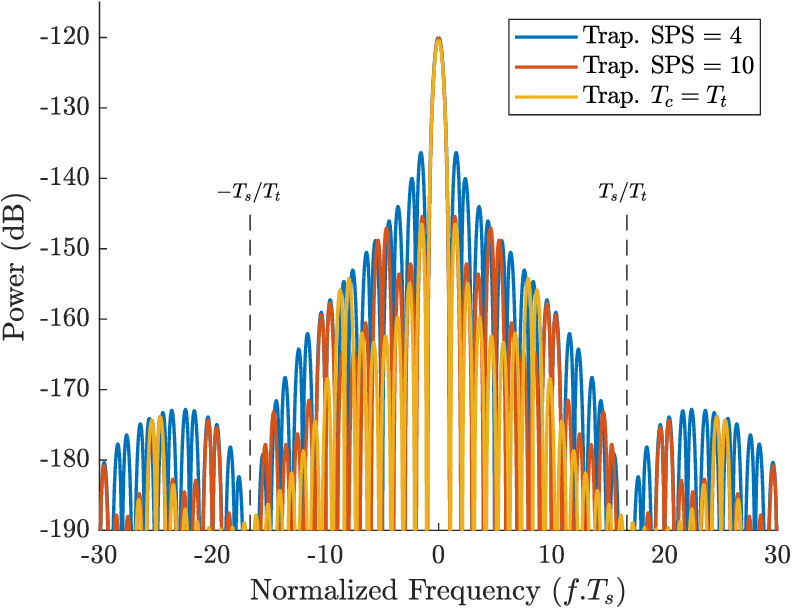}%
} \\ 
\caption{Trapezoidal pulses with hardware impairments and their spectrum.}
\label{fig:steps}
\end{figure}
Spreading the transition of the pulse over a longer duration, such that it is divided into multiple steps, reduces the signal bandwidth. The general expression for an arbitrary pulse divided into multiple samples, each of which has a duration $T_c$ (clock period) can be written as 
\begin{multline}
    x_{\text{sps}}(t) = \sum_{i=0}^{\text{SPS-1}} \frac{a_{i+1}-a_{i}}{T_t}\left[\left(t-iT_c\right)u\left(t-iT_c\right) \right. \\ \left. - \left(t-iT_c-T_t\right)u\left(t-iT_c-T_t\right)\right],
\end{multline}
where $a_i$ presents the final amplitude (after the transition) of each step in the pulse. The Fourier transform of this expression is  
\begin{multline}
    X_{\text{sps}}(f) = \sum_{i=0}^{\text{SPS-1}} \frac{j(a_{i}-a_{i+1})}{2\pi f} sinc \left(T_t \pi f\right) \\ e^{-j2\pi f \left(iT_c+T_t/2\right)}. 
    \label{eq:mini_pulses}
\end{multline}
Compared to Eqs. (\ref{eq:ideal_f}) and (\ref{eq:real_f}), the last equation has the fractional term multiplied by the $sinc$ function inside the summation, which exhibits an amplitude decrease with increasing frequency. This drop can be controlled for each term in the summation by controlling the amplitude change in each step, which is intuitive. 

A trapezoidal pulse formed using this approach and its spectrum are shown in Fig. \ref{fig:steps}. It can be seen that the sidebands of the signal are relatively reduced. However, another problem arises here as a natural consequence of this sampling process due to the periodic term $e^{-j2\pi f \left(iT_c+T_t/2\right)}$ in (\ref{eq:mini_pulses}). While the previous equations have similar periodic components, the harmonics were not clearly seen due to the infinite bandwidth, in contrast to the last approach, where the suppressed sidebands revealed the harmonics clearly. 

Since the SPS is actually increased, baseband pulse shaping filters can be utilized, and their outputs can be applied directly to $a_i$. Before considering the pulse shaping filters, an analysis approach addressing the aforementioned hardware impairments is developed in the next subsection.
\subsection{DFT of Pulses with Higher SPS}
It is shown in (\ref{eq:mini_pulses}) that the spectrum of the pulse modulated using the proposed transmitter can be obtained from the sample amplitude differences. The equation can be written in vector form as
\begin{multline}
    X_{\text{sps}}(f) = \underbrace{\frac{sinc\left(T_t \pi f\right)}{-j2\pi f}e^{-j\pi f T_t}}_{k(f)} \\
    {\underbrace{\begin{pmatrix}
        1 \\ e^{j2\pi f T_c} \\ e^{j4\pi f T_c} \\ e^{j6\pi f T_c} \\ \vdots
    \end{pmatrix}}_{s(f)}}^H
    \underbrace{\begin{pmatrix}
        a_0 - a_1 \\ a_1 - a_2 \\ a_2 - a_3 \\ a_3 - a_4 \\ \vdots
    \end{pmatrix}}_{x'}.
\end{multline}
A Discrete Fourier Transform (DFT) with hardware impairments $(T_t,T_c)$ for this continuous-time signal represented by its amplitude samples ($a_i$) can be obtained by sampling $X_{\text{sps}}(f)$ for different frequencies. The matrix notation for this operation is
\begin{equation}
     X_{\text{sps},T_t,T_c} = \begin{pmatrix}
         k(T_t,f_1) {s(T_c,f_1)}^H \\
         k(T_t,f_2) {s(T_c,f_2)}^H \\ 
         \vdots 
     \end{pmatrix} x'.
\end{equation}
As a result, the spectrum of an arbitrary signal can be calculated directly by taking its sample differences and multiplying them by the formed DFT matrix. The effect of the hardware impairments can then be observed directly. For example, the spectrum of the continuous trapezoidal pulse ($T_c=T_t$) is shown in Fig. \ref{fig:steps} (b). It can be observed from the three plots in Fig. \ref{fig:steps} (b) that reducing the sample period while keeping a constant symbol period (and hence increasing the SPS rate) reduces the sideband levels. 
\subsection{Pulse Shaping and Harmonics}
\begin{figure}[]
\centerline{\includegraphics[width=0.75\linewidth]{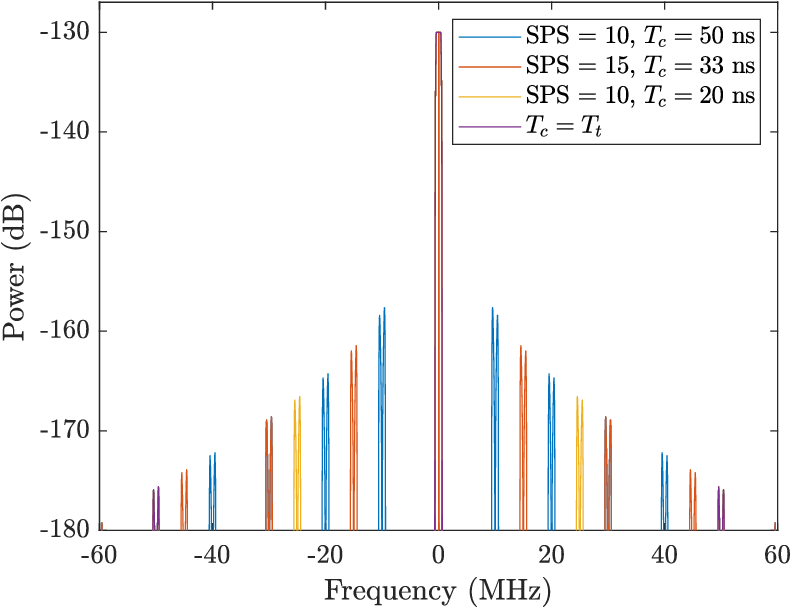}}
\caption{The spectrum of RRC pulse-shaped signals with different
clock periods.}
\label{fig:spectrumsRC}
\end{figure}

While simple rectangular and trapezoidal pulses were considered in the previous sections to develop an analysis model, the widely used root-raised cosine (RRC) pulse shaping is now considered. The frequency in the DFT matrix can take any value. Furthermore, frequency analysis for filter design can be conducted. Here, with the baseband signal being bandlimited, the harmonics from the periodic term in $k(f)$ are clearly seen. A frequency-domain filter design can be implemented by multiplying the filter's frequency response by the inverse DFT matrix. However, it is important to notice that the frequencies in the DFT matrix must satisfy $(-1/T_t \leq f < 1/T_t)$ to obtain a reversible DFT matrix. This periodicity shows that the existence of harmonics is unavoidable by controlling only the signal samples. Thus, the clock period's effect is analyzed as the only controllable design parameter. 
Fig. \ref{fig:spectrumsRC} shows the frequency responses of RRC filter with hardware impairment $T_t=10$ns and a symbol period of $T_s=500$ns. As can be seen from the figure, for a fixed symbol period (fixed bandwidth), reducing the clock period, which is equivalent to increasing the distance between harmonics in the frequency domain, results in reducing the harmonics power significantly. For example, the first harmonic is suppressed by 27 dB, 31 dB, 36 dB, and 45 dB for $T_c=50,33,20,T_t$ ns cases, respectively. It is important to mention here that increasing the signal bandwidth results in increasing the harmonic levels, thus, the key parameter here is the SPS. Higher SPS is required to obtain better sideband suppression, resulting in lower bandwidth for a given transition time $T_t$.

\section{Numerical Analysis and Experimental Results}
\label{Extext}
\begin{figure*}[htbp]
\centerline{\includegraphics[width=0.7\linewidth]{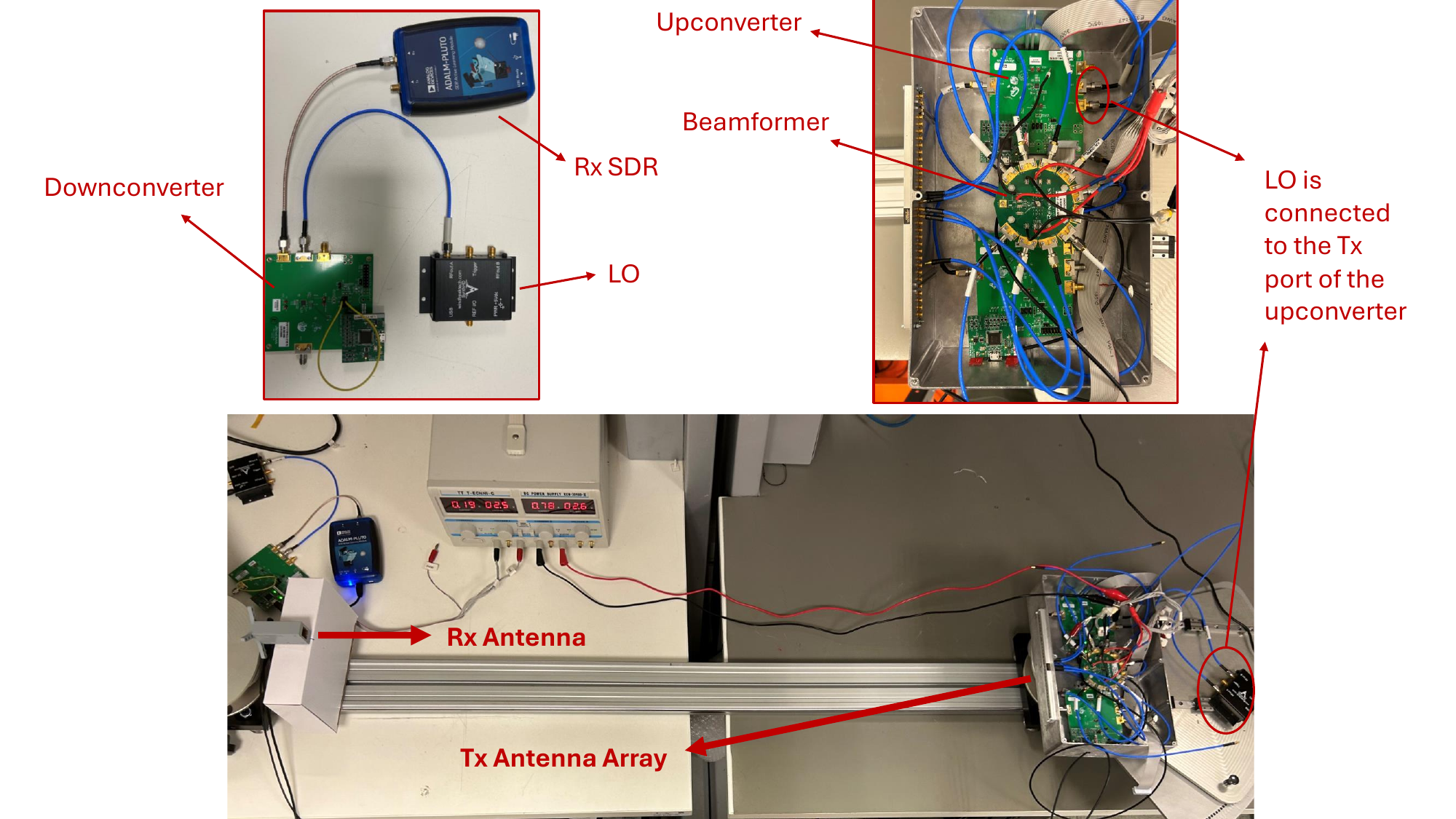}}
\caption{The experimental setup.}
\label{fig:setup}
\vspace{-10pt}
\end{figure*}

Fig. \ref{fig:setup} shows the experimental setup for the proposed transmitter and a software-defined radio (SDR)-based receiver. On the transmitter side, Windfreak SynthHD is used as a signal source to generate two different tones fed to Renesas F5728 upconverter. The first output is a 5.5 GHz tone of -3 dBm power connected to the LO port of the upconverter. The second output generates a 3 GHz tone of -33 dBm connected to the TX port of the upconverter. The upconverter multiplies the LO signal by 4 and mixes it with the TX signal, resulting in an unmodulated carrier at 25 GHz. This carrier is fed to Renesas F5268 analog beamformer, which provides 4 outputs with separated amplitude and phase control. The phase shifters and attenuators are controlled by a 6-bit register for each. The beamformer's outputs are connected to 4 elements of a 32-element antenna array. The receiver is composed of a horn antenna connected to Renesas F5728 downconverter. The LO port of the downconverter is fed with a 5.5 GHz tone generated by another Windfreak SynthHD and the downconverted signal is observed at its RX port around 3 GHz. The downconverted signal is then collected using AdalmPluto SDR. Using the fast beamsteering function in Renesas F5268, the fast update of the phase shifters and attenuators can be realized with a controllable clock rate and a response time $T_t=20$ns \cite{renesas}. The distance between the transmitter and the receiver is 1 meter.

\begin{figure}[htbp]
\centering
\subfloat[QPSK constellation diagram.]{%
\includegraphics[width=0.70\columnwidth]{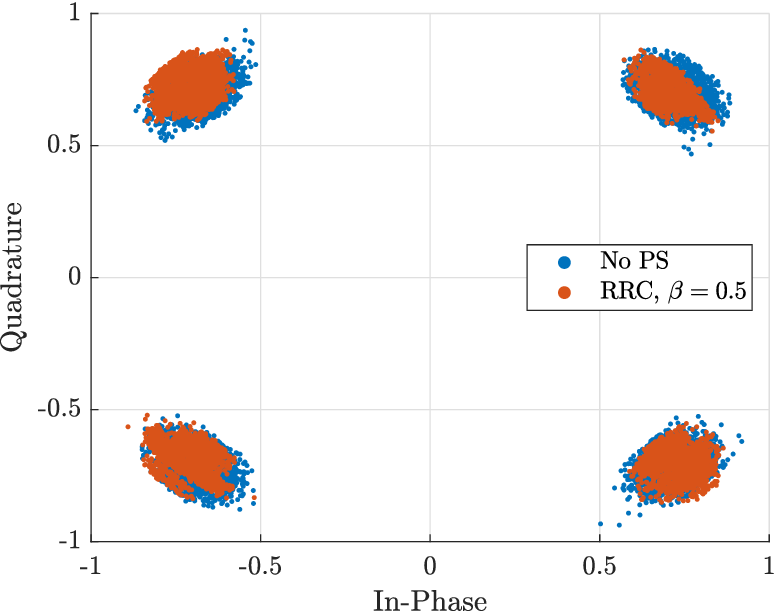}%
} \\
\subfloat[Power spectral density.]{%
\includegraphics[width=0.70\columnwidth]{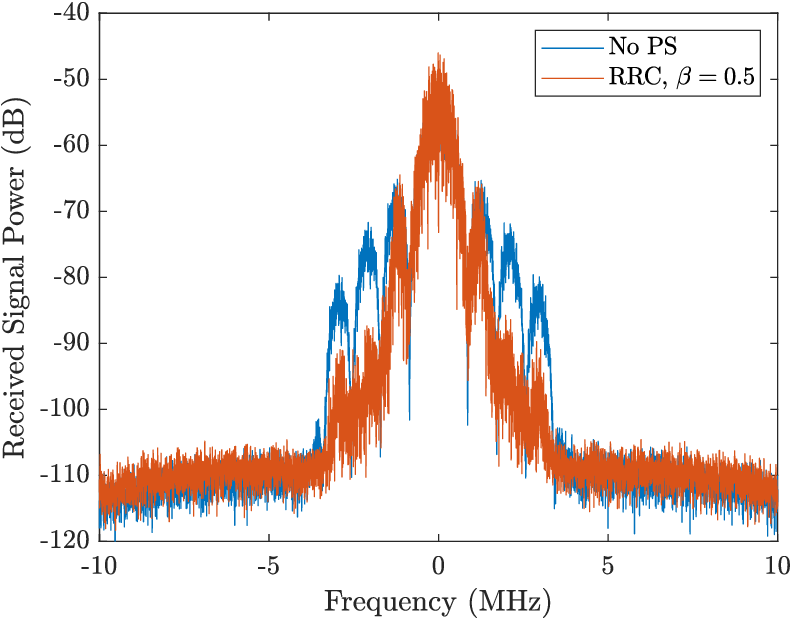}%
} \\ 
\caption{The received signal from a conventional transmitter at 25 GHz with and without pulse shaping.}
\label{fig:deney1_4}
\vspace{-10pt}
\end{figure}

To obtain reference measurements, instead of the 3 GHz single tone, the upconverter is fed by a quadrature phase shift keying (QPSK) signal modulated using another AdalmPluto SDR with a symbol rate of 1 Msps (Mega symbol per second). The results of this experiment are reported for transmission without pulse shaping, and with RRC (roll-off factor $\beta = 0.5$) in Fig. \ref{fig:deney1_4}. The SPS is taken as 8 in both transmissions as well as the receiver for generating the constellation. The bandwidth of the receiver is set to 40 MHz for power spectral density. The power in the figure is in decibels relative to full scale (dBFS). The phase error of the received constellation in this case is measured as 3.13 and 2.49 degrees for the no pulse shaping and RRC cases, respectively. The purpose of this experiment is to provide a reference to compare the proposed transmitter's performance. 

\begin{figure}[htbp]
\centering
\subfloat[QPSK constellation diagram.]{%
\includegraphics[width=0.7\columnwidth]{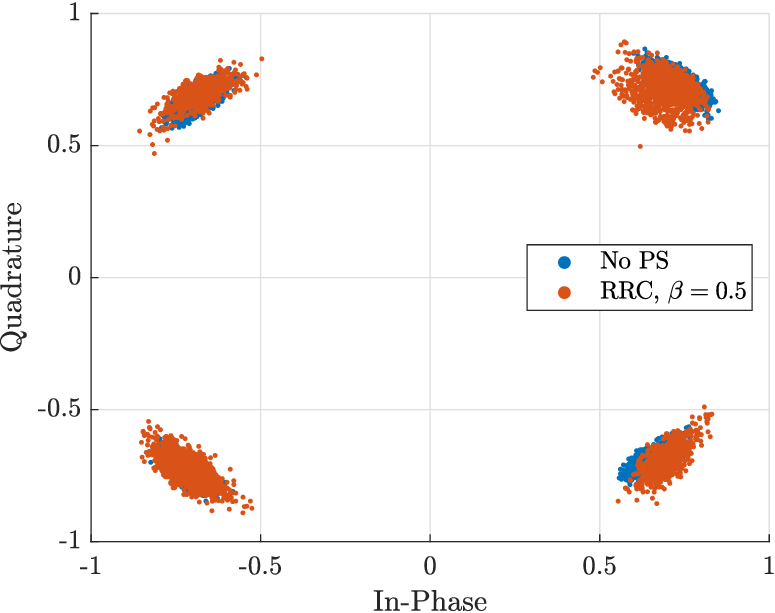}%
} \\
\vspace{-5pt}
\subfloat[Power spectral density.]{%
\includegraphics[width=0.7\columnwidth]{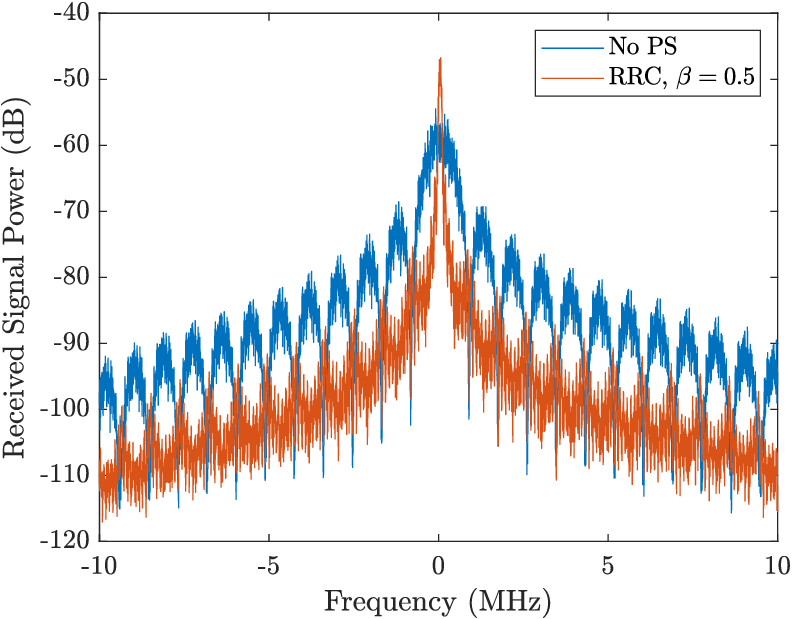}%
} \\ 
\caption{The received signal from the proposed transmitter at 25.1 GHz with and without pulse shaping.}
\label{fig:deney5_12}
\vspace{-10pt}
\end{figure}

The first experiment using the proposed transmitter is to transmit QPSK symbols without pulse shaping. The clock rate of updating the beamformer weights is 857 kHz equivalent to $T_c = 1.16\mu$s and each symbol is presented by one clock (SPS =1). The received signal's spectrum is shown in Fig. \ref{fig:deney5_12} (b), note that the observable received signal's bandwidth is limited by the receiver SDR. The constellation diagram of the received signal is shown in Fig. \ref{fig:deney5_12} (a) and the phase error is measured as 2.11 degrees. 

The second experiment is to apply pulse shaping. With the same clock rate and an SPS of 8, the baseband signal bandwidth is limited to 107 kHz, and the harmonics are seen in Fig. \ref{fig:deney5_12} (b) at the multiplications of 857 kHz. The first harmonic is measured at 29 dB below the original signal's power. The constellation diagram of this signal is given in Fig. \ref{fig:deney5_12} (a) and the phase error is measured as 2.99 degrees. The reason for this increase in the phase error is the quantization error due to the 6-bit register control in the phase shifters and attenuator. While the direct modulation of QPSK has only 4 distinct values, the pulse-shaped signal has much more sample values where the quantization effect arises. 

\begin{figure}[htbp]
\centerline{\includegraphics[width=0.7\linewidth]{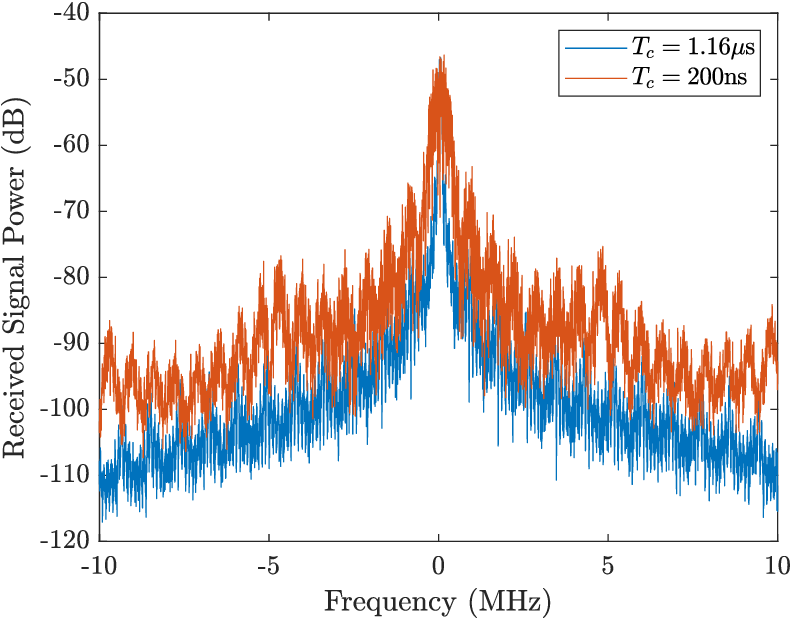}}
\caption{The power spectral density of signals with different clock rates and fixed SPS.}
\label{fig:deney13_15}
\vspace{-13pt}
\end{figure}

In the third experiment, the clock rate was swept over different values with fixed SPS RRC-shaped signals. Since the SPS is not changed, the signal's bandwidth increases as the clock rate increases, which results in less harmonic suppression. For example, the power spectral density of signals with two different clock rates is shown in Fig. \ref{fig:deney13_15}. The measurement results show that the first harmonic suppression is $29$ dB for 5 different clock rates, each one corresponding to a different signal bandwidth. 
\section{Conclusion}
\label{conclusiontext}
Generating and transmitting signals at mmWave frequencies suffers from high cost and phase noise challenges. This paper investigates modulating baseband signal samples directly to the antenna elements using analog beamformers. The theoretical analysis of this transmission scheme is conducted and it is shown that digital pulse shaping can be applied and the harmonics resulting from the sampling process can be reduced by reducing the sampling period and increasing the SPS rate. An experimental demonstration of the proposed transmission model using Renesas F5268 beamformers has been shown and the results were consistent with theoretical conclusions. In future work, different waveforms with higher bandwidths will be analyzed, and solutions for the beamformer quantization errors will be developed.
\vspace{-5pt}
\section{Acknowledgment}
This study has been carried out through the research vision of the THULAB project run at the Informatics and Information Security Research Center (B{\.I}LGEM) of The Scientific and Technological Research Council of Türkiye (T{\"U}B{\.I}TAK).

This work has also received funding from the EcoMobility  project. EcoMobility has been accepted for funding within the CHIPS Joint Undertaking, a public-private partnership in collaboration with the HORIZON Framework Programme and the National Authorities under grant agreement number 101112306.

\vspace{-5pt}
\bibliographystyle{IEEEtran}

\bibliography{refs}

\end{document}